\documentclass[proceedings]{JHEP3}

\PrHEP{PrHEP hep2001}
\conference{International Europhysics Conference on HEP}               
\usepackage{epsfig}  
\title{\Large Physics Results from the AMANDA Neutrino Detector}

\author{\speaker{Marek Kowalski} for the AMANDA collaboration \\
        Address: DESY-Zeuthen, Platanenallee 6, D-15738 Zeuthen, Germany\\ 
        E-mail: \email{marek.kowalski@desy.de}
}
\author{

{\small
The AMANDA Collaboration:
J.~Ahrens$^{9}$, 
X.~Bai$^{1}$, 
G.~Barouch$^{12}$, 
S.W.~Barwick$^{8}$, 
R.C.~Bay$^{7}$, 
T.~Becka$^{9}$, 
K.-H.~Becker$^{2}$, 
D.~Bertrand$^{3}$, 
A.~Biron$^{4}$, 
S.~B\"oser$^{4}$
% ...
}

\vspace{0.2cm}
{\scriptsize

   (1) Bartol Research Institute, University of Delaware, Newark, DE 19716, USA\newline
   (2) Fachbereich 8 Physik, BUGH Wuppertal, D-42097 Wuppertal, Germany 
\newline
   (3) Univ. Libre de Bruxelles, Science Faculty CP230, Boulevard du Triomphe, B-1050 Brussels, Belgium 
\newline
   (4) DESY-Zeuthen, D-15735 Zeuthen, Germany
\newline
   (5) Dept. of Technology, Kalmar University, S-39182 Kalmar, Sweden
\newline
   (6) Lawrence Berkeley National Laboratory, Berkeley, CA 94720, USA
\newline
   (7) Dept. of Physics, University of California, Berkeley, CA 94720, USA
\newline
   (8) Dept. of Physics and Astronomy, University of California, Irvine, CA 92697, USA
\newline
   (9) Institute of Physics, University of Mainz, Staudinger Weg 7, D-55099 Mainz, Germany
\newline
   (10) University of Mons-Hainaut, 7000 Mons, Belgium
\newline
   (11) Dept. of Physics and Astronomy, University of Pennsylvania, Philadelphia, PA 19104, USA
\newline
   (12) Dept. of Physics, University of Wisconsin, Madison, WI 53706, USA
\newline
   (13) Physics Department, University of Wisconsin, River Falls, WI 54022, USA
\newline
   (14) Division of High Energy Physics, Uppsala University, S-75121 Uppsala, Sweden
\newline
   (15) Fysikum, Stockholm University, S-11385 Stockholm, Sweden
\newline
   (16) Vrije Universiteit Brussel, Dienst ELEM, B-1050 Brussel, Belgium
\newline
   (17) Present address: CERN, CH-1211, Gen\`eve 23, Switzerland

}}

\abstract{In the winter season of 2000, the AMANDA (Antarctic Muon And
Neutrino Detector Array) detector was completed to its final state. We
report on major physics results obtained from the AMANDA-B10 detector, 
as well as initial results of the full AMANDA-II detector.}

\begin{document}

\section{Introduction}

The AMANDA-II high energy neutrino detector was constructed 1500-2000 m below
the surface of the Antarctic ice sheet at the geographic South Pole.
In the austral summer 1999/2000, the detector was
completed to its final state, consisting of
677 Optical Modules (OMs) on 19 strings.  
Figure \ref{fig:eiffel} shows a schematic view of the AMANDA-II detector.
The OMs consist of 8"
 Photo Multiplier Tubes (PMTs) housed by pressure resistant glass spheres.
The OMs on the newer 
 strings 11-19  are connected via fiber-optic cables for high
bandwidth transmission of the analog PMT signal.
Strings 1-10, deployed
in the seasons of 1995/1996 and 1996/1997, use electric cable transmission.
\EPSFIGURE[t]{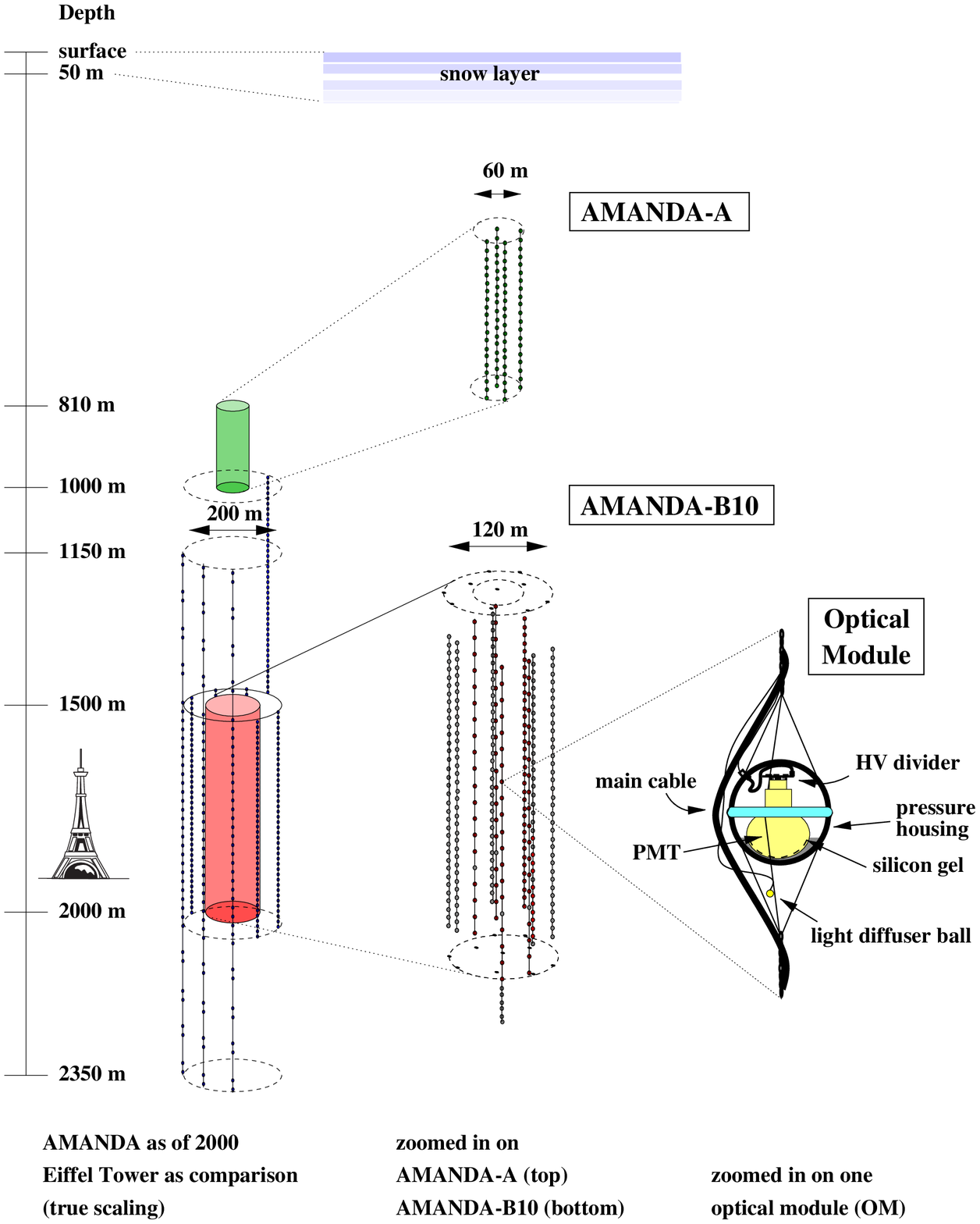,width=6.9cm}
{The AMANDA-II detector, as of 2000.}
\label{fig:eiffel}

The main channel for detecting high energy neutrinos is through
observation of the muons from charged current,
$\nu_\mu$-Nucleon interactions.
The \v{C}erenkov light from a highly energetic muon is picked up by the
PMTs, and is used for track reconstruction. Besides the desired 
astrophysical neutrinos, there is a significant background of atmospheric muons
and neutrinos, produced by cosmic rays interacting
in the atmosphere. The observation of atmospheric neutrinos is a
crucial performance test for the detection capabilities of AMANDA, as
they represent a known flux of neutrinos. Here, we  report on the observation 
of atmospheric neutrinos in the data taken in 1997. We further summarize 
the searches for astrophysical neutrinos, within the same data.
We then give a brief overview of the present stage of analysis of the data
taken with the fully completed AMANDA-II detector.

\section{Atmospheric Neutrinos}
The main challenge in the investigation of atmospheric neutrinos is the
large background of atmospheric muons \cite{andres01}.  
While there are about $8.5\cdot 10^6$ muons a day, with
energies high enough ($>$30-50 GeV) to 
trigger the detector, we expect only a few tens of muons from
atmospheric neutrinos. The muons from neutrinos are identified by
having upward reconstructed tracks. The reconstruction of the muon
track is done with a maximum likelihood method, which incorporates both
scattering and absorption of photons in the ice \cite{wiebusch99}.
\EPSFIGURE{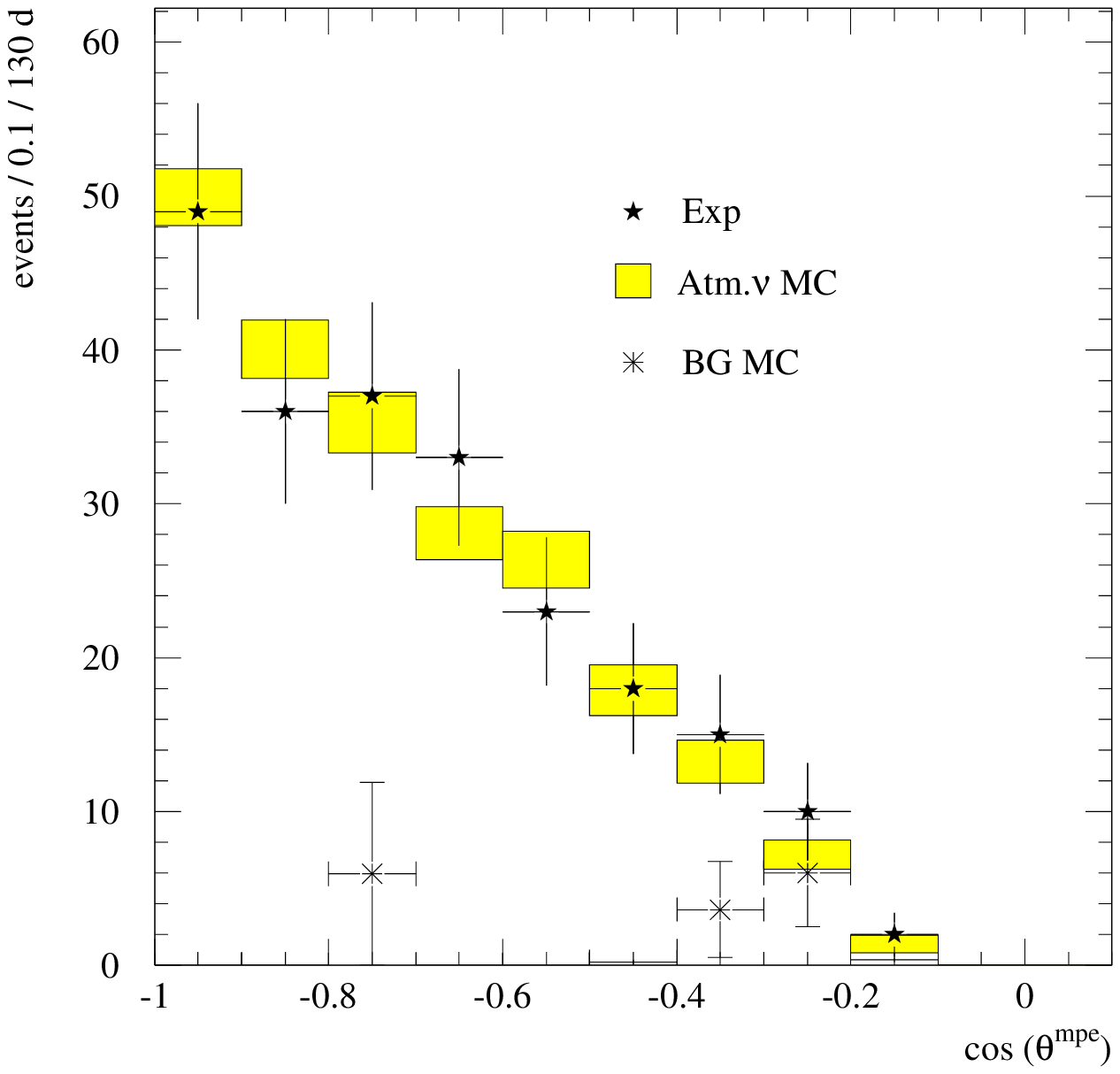,width=7.4cm}   
{Distribution of the cosine of the reconstructed zenith angle, after full application of the neutrino selection cuts.
Shown are the experimental data points with the MC expectations for the 
atmospheric $\mu$-background and the atmospheric neutrino simulation. Note that the BG simulation is  limited by statistics. The error bars shown include only the statistical error. 
Additionally there is a factor of 2 uncertainty in the overall normalization of the atmospheric neutrino MC.}
\label{fig:zenith97}
Two working groups performed two independent analyses
 of the same set of data, leading to comparable and consistent results
 \cite{andres01, wiebusch01, deyoung01}.  
Each of the two analyses finds about 200 neutrino events in 130 days of
detector live-time.

Combining the samples, there are 325 neutrino candidates.
Figure \ref{fig:zenith97} shows for one of the analysis the distribution of the cosine of the reconstructed zenith angle for experimental data, atmospheric 
neutrino Monte Carlo (MC), as well as background muon MC.
The neutrino MC includes neutrino oscillations with 
values sin$^22\theta$=1 and $\Delta m^2=0.003$~eV$^{2}$
leading to a 10~\% reduction of the event rate.
The  celestial distribution of events is consistent  with 
a random, isotropic distribution.   
Different methods lead 
to an estimation of about 10\% background contamination in the final sample.
The observed rates are between $\sim$ 0\% and 50\%  lower than
predicted by MC, where the uncertainties are due to theoretical
uncertainties in the neutrino flux and due to various experimental
uncertainties, mainly the precise knowledge of the properties of the
ice in the close surrounding of the Optical Modules. 
For a full discussion of the uncertainties of the signal and
background MC see \cite{wiebusch01}. 

\section{Search for a diffuse high energy neutrino flux}
The search for a diffuse neutrino flux of astrophysical  origin
follows naturally from the measurement of the diffuse flux of
atmospheric neutrinos. Neutrinos from generic astrophysical sources
are expected to extend to higher energies while the energy spectrum of
atmospheric neutrinos falls of steeply with increasing energy. 
A simple and robust measure of the energy of the observed muons is the
number of PMTs that detected at least one photoelectron in a given
event. Figure \ref{fig:nch} shows the distribution of the number of fired PMTs
for the observed experimental data, atmospheric neutrino MC and a hypothetical flux of astrophysical neutrinos following a
power law of $ E_\nu^{2}dN/dE_{\nu}=10^{-5}$cm$^{-2}$s$^{-1}$sr$^{-1}$GeV. 
The assumed flux of astrophysical neutrinos would generate
an excess at high multiplicities of fired PMTs. 
The analysis
\cite{hill01,leu01} does not show such an excess.
This leads to a preliminary limit (at 90~\%C.L.) on the ($E^{-2}$) flux of high
energy neutrinos of 
$E_\nu^{2}dN/dE_{\nu}=0.9 \cdot 10^{-6}$cm$^{-2}$s$^{-1}$sr$^{-1}$GeV. 
This limit is in the range of different predictions for the flux of neutrinos
from Active Galactic Nuclei \cite{AGN_predictions,Nellen:1993dw} as well as generic theoretical bounds for the diffuse flux of astrophysical neutrinos
\cite{Mannheim00}. 

The AMANDA limit is below 
previously stated experimental limits from  FREJUS \cite{frejus96}, MACRO \cite{macro01} and BAIKAL \cite{baikal01}.
AMANDA is also searching for a diffuse 
flux of $\nu_{e}$ neutrinos, which
appear as electromagnetic and hadronic showers, producing detectable 
bursts of light in the close vicinity of the detector \cite{taboada01}.
\EPSFIGURE{nch_bw.epsi,width=7.cm}   
{Shown is the distribution of the number of fired OMs for experimental data, MC expectation of atmospheric neutrinos and a hypothetical flux of astrophysical neutrinos (AGN). 
}
\label{fig:nch}

\section{Search for point sources}
If the dominant astrophysical neutrino flux is emitted by a few particular
bright or close sources, their direction could be resolved. For this reason,
the data taken with the AMANDA-B10 detector was analyzed for neutrinos from
 point sources \cite{scott01}. 
The sky was subdivided into 154 search bins, each with a half width of about 
$5^\circ$. Due to the lower background within each search bin, it is
possible to relax the cuts and thus gain a higher signal efficiency. The
probability for the observed occupations of the bins shows no
significant deviation from a pure background distribution. A preliminary 
limit on the muon flux from neutrinos from point sources is given 
(again assuming an
$E_\nu^{-2}$ spectrum):\newline  
$\Phi(E_\mu>10~$GeV$)= 0.6 - 7 \cdot 10^{-13} $cm$^{-2}$s$^{-1}$, 
where the lower and higher values of the limit corresponds to the vertical 
and horizontal direction. The sensitivity for point sources reached by
AMANDA-B10 is only about one
order of magnitude above the observation of TeV gamma ray fluxes from  the
blazar Mrk 501 during its flaring phase in 1997.
\section{Some more Results from AMANDA-B10}
Other results from AMANDA-B10 include a limit on the flux of neutrinos from
 Gamma Ray Bursters observed by the BATSE satellite \cite{rellen01}, a limit on
 the flux of  relativistic magnetic monopoles \cite{niessen01} and a limit on
 the neutralino annihilation rate within the earth~\cite{cph01}. Further,
 AMANDA is monitoring our Galaxy for supernovae explosions, by observing PMT 
noise rate variations \cite{bouchta01}.  
\EPSFIGURE[t]{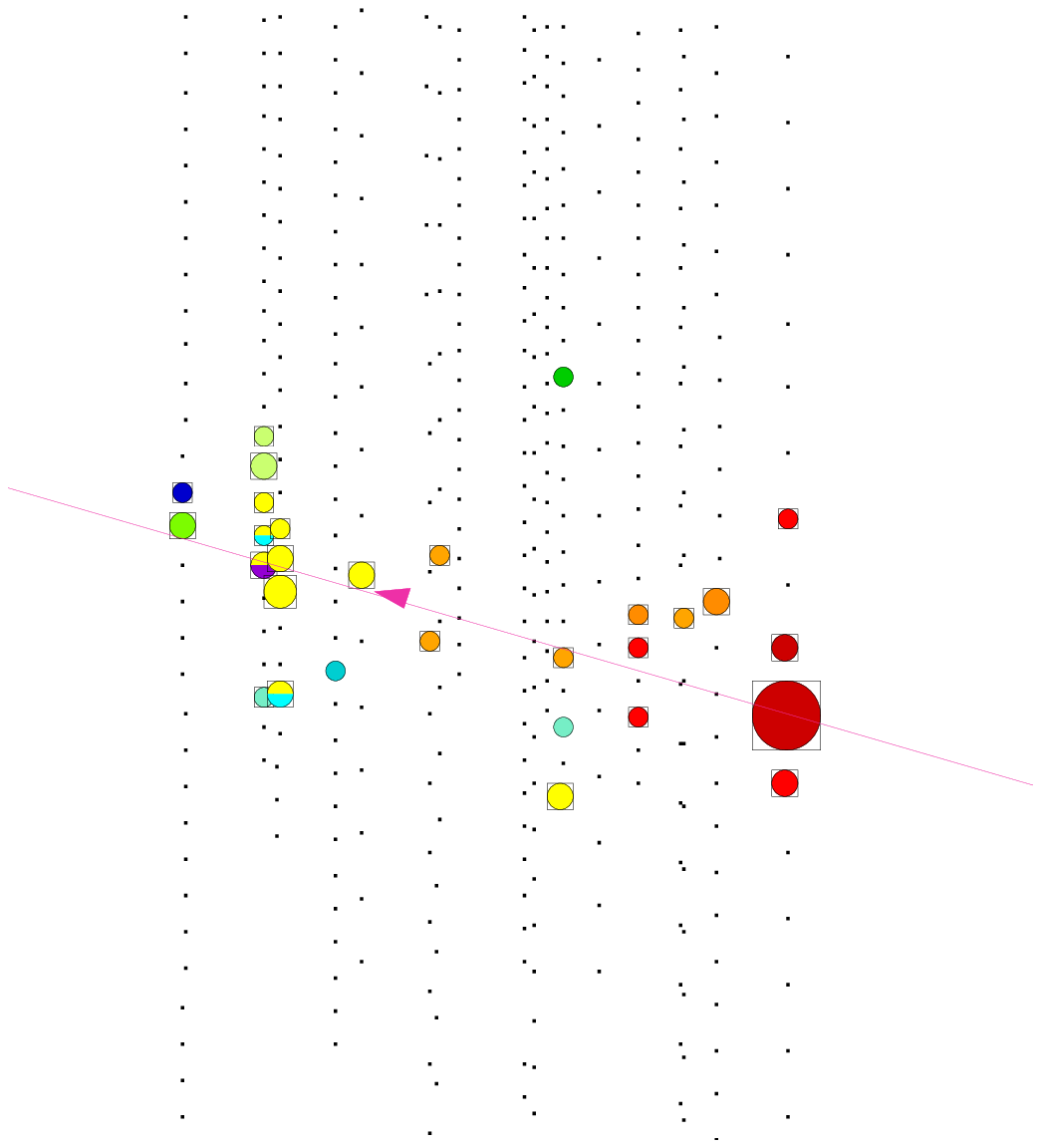,width=7cm}
{Event display of a neutrino candidate from the AMANDA-II 2000  data. 
The colored circles indicate fired PMTs (red means early, blue late). 
The line represents the reconstructed track. Only a horizontal slice of the detector is shown.}
\label{fig:event}
\section{The completed AMANDA-II detector}
The completed AMANDA-II detector was calibrated during the austral summer of 
2000/2001. Since then, 20~\% of the data has been analyzed.
With its additional outer ring of 9 strings, AMANDA-II has 
considerably improved muon track reconstruction capabilities.
This is in particular true for tracks close to the horizon. 
Compared to AMANDA-B10, this leads to an enlarged angular acceptance \cite{wischnew01}. The fraction of signal MC retained from the trigger level for the
preliminary atmospheric neutrino analysis increased from about 5~\% for 

\noindent
AMANDA-B10 to 20~\% for AMANDA-II. With a higher energy threshold due to a
stricter trigger setting, we observe about 4 neutrino candidates per day
with AMANDA-II.
From the beginning of this year, an online reconstruction and filter has been installed. On a daily basis, we obtain neutrinos candidates similar in rate to 
that of the above described offline analysis. 

With 2-3 years of AMANDA-II data, we expect to gain 0.5-2 orders of 
magnitude in sensitivity for various astrophysical neutrino sources 
\cite{barwick01}. 
The next big step in improving the sensitivity to astrophysical neutrino sources will be the construction of the IceCube-detector \cite{cspier01},     
a telescope with a 1~km$^3$ instrumented volume and 5000 OMs on 80 strings
surrounding AMANDA-II.


\end{document}